\documentclass[aps,prd,preprint,showpacs,groupaddress]{revtex4}
\usepackage{color}
\usepackage{graphicx}
\usepackage{tabularx}

\begin{document}

\title{The Dark Matter Constraints on the Left-Right Symmetric Model with $Z
_2$ Symmetry}

\author{\footnotesize Wan-lei Guo\footnote{guowl@itp.ac.cn}, Li-ming Wang\footnote{wanglm@itp.ac.cn},
 Yue-liang Wu\footnote{ylwu@itp.ac.cn}, Ci Zhuang\footnote{zhuangc@itp.ac.cn} }

\address{Kavli Institute for Theoretical Physics China,
Institute of Theoretical Physics, \\ Chinese Academy of Science,
Beijing 100080, P.R.China}

\begin{abstract}

In the framework of Left-Right symmetric model, we investigate an
interesting scenario, in which the so-called VEV seesaw problem can
be naturally solved with $\cal{Z}$$_2$ symmetry. In such a scenario,
we find a pair of stable weakly interacting massive particles
(WIMPs), which may be the cold dark matter candidates. However, the
WIMP-nucleon cross section is 3-5 orders of magnitude above the
present upper bounds from the direct dark matter detection
experiments for $m \sim 10^2-10^4 $ GeV. As a result, the relic
number density of two stable particles has to be strongly suppressed
to a very small level. Nevertheless, our analysis shows that this
scenario can't provide very large annihilation cross sections so as
to give the desired relic abundance except for the resonance case.
Only for the case if the rotation curves of disk galaxies are
explained by the Modified Newtonian Dynamics (MOND), the stable
WIMPs could be as the candidates of cold dark matter.

\end{abstract}


\maketitle

\section{Introduction}

The Left-Right (LR) symmetric model \cite{LR}, based on the gauge
group $SU(2)_L \times SU(2)_R \times U(1)_{B-L}$, is an attractive
extension of the standard model (SM). The symmetry requires the
introduction of right-handed partners for the observed gauge bosons
and neutrinos, and a Higgs sector containing one bi-doublet $\phi$
(2,2,0), one left-handed triplet $\Delta_L$ (3,1,2) and one
right-handed triplet $\Delta_R$ (1,3,2). In such a minimal LR
symmetric model, parity is an exact symmetry of the theory at high
energy scale, and is broken spontaneously at low energy scale due to
the asymmetric vacuum. Also CP asymmetry can be realized as a
consequence of spontaneous symmetry breaking, namely the spontaneous
CP violation (SCPV) \cite{SCPV}. However, such a scenario suffers
from nontrivial constraints from the vacuum minimization conditions.
It is explicitly demonstrated that the SCPV is not so easily
realized if all the parameters in the Higgs potential are real and
endowed with natural values \cite{Mohapatra,VEV,Barenboim}. The
difficulty results from the facts that one of the neutral Higgs
bosons carries dangerous tree level flavor changing neutral currents
(FCNC) effect, and that quark flavor mixing angles and CP violating
phase are all calculable quantities due to the LR symmetry.
Therefore, many generalized CP violation scenarios beyond the SCPV
case have been analyzed extensively \cite{Langacker,Bernabeu,Pos,
Frere, Kie,Ji}. In these literatures, the masses of right-handed
gauge boson $W_2$ and the FCNC Higgs boson are strongly constrained
from low energy phenomenology. Although the CKM matrix are more
general not to be fully fixed than the SCPV case, it is proved that
there is only one physical complex phase in the Yukawa couplings
\cite{Ken}. Hence the FCNC Higgs boson's couplings can't be
absolutely free. The FCNC Higgs boson's mass still accepts strict
bound. In terms of these observations, a generalized two Higgs
bi-doublets model is proposed \cite{Wu}. In this model, quark mass
matrices become far more flexible and the FCNC Higgs boson's Yukawa
couplings are now free parameters. Thereby low energy bound on the
right-handed scale is largely alleviated. As other generalized
models, the two Higgs bi-doublets version of LR model also has the
advantage to realize the SCPV without the fine-tuning problem.

The LR symmetric model is also motivated to explain the very tiny
neutrino masses. When the vacuum expectation value (VEV) $v_R$ of
the neutral component of $\Delta_R$ is very huge, typically of order
$10^{12}$ GeV, the well-known seesaw mechanism provides a very
natural explanation of the smallness of neutrino masses
\cite{SEESAW}. However, the right-handed gauge bosons $Z_2$ and
$W_2$ are too heavy to be detected at the Large Hadron Collider
(LHC) and the future colliders. To allow for the possibility of an
observable right-handed scale, many authors focus on the $v_R \sim
10 $ TeV case. Although the seesaw mechanism can work well, we have
to face the so-called VEV-seesaw puzzle. Namely, $\beta/\rho$ is of
order $10^{-10}$ rather than the anticipant ${\cal O} (1)$, where
$\rho$ and $\beta$ are located in the Higgs potential. One may
introduce a discrete ${\cal Z}_2$ symmetry $\Delta_L \rightarrow -
\Delta_L$ and $\Delta_R \rightarrow  \Delta_R$ to resolve this
VEV-seesaw problem \cite{VEV}. It is worthwhile to stress that
neutrinos are the Dirac particles in this scenario. If we preserve
the Majorana Yukawa couplings, the corresponding model must lie
beyond the LR symmetric model.

The ${\cal Z}_2$ symmetry leads to the absence of both $\beta$-type
terms and the Majorana Yukawa couplings, hence $v_L = 0$ due to the
minimization conditions. Furthermore, we find that the neutral Higgs
bosons $\delta_L^0$ and ${\delta_L^0}^*$ are a pair of stable weakly
interacting massive particles (WIMPs). This is an important feature
of our scenario which hasn't been indicated before.  It is a natural
idea that $\delta_L^0$ and ${\delta_L^0}^*$ may be the cold dark
matter candidates \cite{DM}. We firstly calculate the WIMP-nucleon
elastic scattering cross section which has been strongly constrained
by the direct dark matter detection experiments, such as the
CDMS\cite{CDMS} and XENON\cite{XENON}. However, our result is 3-5
orders of magnitude above the present bounds  for $m \sim 10^2-10^4
$ GeV \cite{CDMS, XENON}. To avoid this puzzle, $\delta_L^0$ and
${\delta_L^0}^*$ can't dominate all the dark matter. We find that
our scenario is consistent with the direct dark matter detection
experiments only when $n_{\delta_L^0} \leq 4.8 \times 10^{-14}$,
where $n_{\delta_L^0}$ is the total relic number density of
$\delta_L^0$ and ${\delta_L^0}^*$. This bound requires the dark
matter annihilation cross sections must be very large. In this work,
we examine whether our scenario can provide very large annihilation
cross sections so as to derive the desired relic abundance.

In this paper we try to give a comprehensive analysis on these LR
models with general parameter setting. Firstly, we perform a
detailed investigation on the simplest LR model with one Higgs
bi-doublet, in which there are no any CP violation phases. Then we
generalize the simplest LR model to some other more complicated
situations. It turns out that there's no significant differences
among these one Higgs bi-doublet versions of LR model because the
gauge and Higgs sectors are basically the same. Whereas in the two
Higgs bi-doublet case, there would be more Higgs bosons and the
Yukawa couplings might be quite different. Hence more delicate
analysis is needed. The remaining part of this paper is organized as
follows. In Section II, we briefly describe the main features of the
LR symmetric model and discuss the VEV-seesaw problem. In Section
III and IV, the direct dark matter detection experiments put very
strong constraints on the relic number density and the annihilation
cross sections. In Section V, we analyze whether the simplest LR
model can be consistent with the above constraints or not.  Then we
generalize the simplest LR model to the two Higgs bi-doublets case
in Sec VI. The summary and comments are given in Section VII.

\section{The LR symmetric model with ${\cal Z}_2$ symmetry}

The minimal LR symmetric model consists of one Higgs bi-doublet
$\phi$ (2,2,0), one left-handed Higgs triplet $\Delta_L$ (3,1,2) and
one right-handed Higgs triplet $\Delta_R$ (1,3,2), which can be
written as
\begin{eqnarray}
\phi  = \left ( \matrix{ \phi_1^0 & \phi_1^+ \cr \phi_2^- & \phi_2^0
\cr  } \right ) ; \; \Delta_{L,R}  = \left ( \matrix{
\delta_{L,R}^+/\sqrt{2} & \delta_{L,R}^{++} \cr \delta_{L,R}^{0} &
-\delta_{L,R}^{+}/\sqrt{2} \cr } \right ) \;.
\end{eqnarray}
After the spontaneous symmetry breaking, the Higgs multiplets can
have the following vacuum expectation values
\begin{eqnarray}
\langle \phi \rangle  = \left ( \matrix{ \kappa_1/\sqrt{2} & 0 \cr 0
& \kappa_2/\sqrt{2} \cr } \right ) ; \; \langle \Delta_{L,R} \rangle
= \left ( \matrix{0 & 0 \cr v_{L,R}/\sqrt{2} & 0 \cr } \right ) \;,
\end{eqnarray}
where $\kappa_1$, $\kappa_2$, $v_L$ and  $v_R$ are in general
complex. Without loss of generality, one can choose $\kappa_1$ and
$v_R$ to be real, while assign complex phases $\theta_2$ and
$\theta_L$ for $k_2$ and $v_L$, respectively. Following the
requirements of the LR symmetry, we can write down the most general
form of the Higgs potential \cite{VEV}
\begin{eqnarray}
V= &-&\mu_1^2\left({\rm Tr}\left[\phi^{\dagger}\phi \right] \right)
-
\mu_2^2\left({\rm Tr} \left[\tilde{\phi}\phi^{\dagger} \right] + {\rm Tr}\left[\tilde{\phi%
}^{\dagger} \phi \right] \right) -\mu_3^2 \left( {\rm Tr} \left[
\Delta_L \Delta_L^{\dagger} \right] + {\rm Tr} \left[
\Delta_R \Delta_R^{\dagger} \right] \right)  \nonumber \\
&+&\lambda_1 \left( \left( {\rm Tr} \left[ \phi \phi^{\dagger}
\right] \right)^2 \right)+ \lambda_2 \left( \left( {\rm Tr} \left[
\tilde{\phi} \phi^{\dagger} \right] \right)^2 + \left( {\rm Tr}
\left[ \tilde{\phi}^{\dagger} \phi \right] \right)^2 \right) +
\lambda_3 \left( {\rm Tr} \left[ \tilde{\phi} \phi^{\dagger} \right]
{\rm Tr} \left[
\tilde{\phi}^{\dagger} \phi \right] \right)  \nonumber \\
&+& \lambda_4 \left( {\rm Tr} \left[ \phi \phi^{\dagger} \right]
\left( {\rm Tr} \left[ \tilde{\phi} \phi^{\dagger} \right] + {\rm
Tr} \left[ \tilde{\phi}^{\dagger} \phi \right] \right) \right)  +
\rho_1 \left( \left( {\rm Tr} \left[ \Delta_L \Delta_L^{\dagger}
\right] \right)^2+\left( {\rm Tr} \left[ \Delta_R \Delta_R^{\dagger}
\right] \right)^2
\right)  \nonumber \\
&+&\rho_2 \left( {\rm Tr} \left[ \Delta_L \Delta_L \right] {\rm Tr}
\left[ \Delta_L^{\dagger} \Delta_L^{\dagger} \right] + {\rm Tr}
\left[ \Delta_R \Delta_R \right]{\rm Tr} \left[ \Delta_R^{\dagger}
\Delta_R^{\dagger} \right] \right) + \rho_3 \left( {\rm Tr} \left[
\Delta_L \Delta_L^{\dagger} \right] {\rm Tr} \left[
\Delta_R \Delta_R^{\dagger} \right] \right)  \nonumber \\
&+&\rho_4 \left( {\rm Tr} \left[ \Delta_L \Delta_L \right] {\rm Tr}
\left[ \Delta_R^{\dagger} \Delta_R^{\dagger} \right] + {\rm Tr}
\left[ \Delta_L^{\dagger} \Delta_L^{\dagger} \right] {\rm Tr} \left[
\Delta_R \Delta_R \right] \right)
\nonumber \\
&+& \alpha_1 \left( {\rm Tr} \left[ \phi \phi^{\dagger} \right]
\left( {\rm Tr} \left[ \Delta_L \Delta_L^{\dagger} \right] + {\rm
Tr} \left[ \Delta_R \Delta_R^{\dagger} \right] \right) \right)
\nonumber \\
&+& \alpha_2 \left( {\rm Tr} \left[ \phi \tilde{\phi}^{\dagger}
\right] {\rm Tr} \left[ \Delta_R \Delta_R^{\dagger} \right]+ {\rm
Tr} \left[ \phi^{\dagger} \tilde{\phi}
\right] {\rm Tr} \left[ \Delta_L \Delta_L^{\dagger} \right] \right)  \nonumber \\
&+&\alpha^\ast_2 \left( {\rm Tr} \left[ \phi^{\dagger} \tilde{\phi}
\right] {\rm Tr} \left[ \Delta_R \Delta_R^{\dagger} \right]+ {\rm
Tr} \left[ \tilde{\phi}^{\dagger} \phi \right] {\rm Tr} \left[
\Delta_L \Delta_L^{\dagger} \right] \right) \nonumber \\
&+& \alpha_3 \left( {\rm Tr} \left[ \phi \phi^{\dagger} \Delta_L
\Delta_L^{\dagger} \right] +{\rm Tr} \left[ \phi^{\dagger} \phi
\Delta_R
\Delta_R^{\dagger} \right] \right)  \nonumber \\
&+& \beta_1 \left( {\rm Tr} \left[ \phi \Delta_R \phi^{\dagger}
\Delta_L^{\dagger} \right] +{\rm Tr} \left[ \phi^{\dagger} \Delta_L
\phi \Delta_R^{\dagger} \right] \right)  + \beta_2 \left( {\rm Tr}
\left[ \tilde{\phi} \Delta_R \phi^{\dagger} \Delta_L^{\dagger}
\right] +{\rm Tr} \left[ {\tilde{\phi}}^{\dagger} \Delta_L \phi
\Delta_R^{\dagger} \right] \right)  \nonumber \\
&+& \beta_3 \left( {\rm Tr} \left[ \phi \Delta_R
{\tilde{\phi}}^{\dagger} \Delta_L^{\dagger} \right] + {\rm Tr}
\left[ \phi^{\dagger} \Delta_L \tilde{\phi} \Delta_R^{\dagger}
\right] \right)\; ,
\end{eqnarray}
where $\tilde{\phi} = \tau_2 \phi^* \tau_2$ and all parameters
$\mu_i$, $\lambda_i$, $\rho_i$, $\alpha_i$ and $\beta_i$ are real.
Only $\alpha_2$ can be complex.  The phases of $\kappa_2$ and $v_L$
may lead to the SCPV \cite{SCPV}. It has been shown that the combing
constraints from $K$ and $B$ system actually exclude the minimal LR
symmetric Model with the SCPV in the decoupling limit \cite{Frere}.
For our present purpose, we investigate here the simplest LR model,
in which $\alpha_2$, $\kappa_2$, $v_L$ and the Yukawa couplings are
real. It is worthwhile to stress that our remaining analysis can be
generalized to the other CP violation scenarios
\cite{Langacker,Bernabeu,Pos, Frere, Kie,Ji}.

In the minimal LR symmetric model, the Lagrangian relevant for the
neutrino masses reads \cite{VEV}:
\begin{eqnarray}
-{\cal L}  =   Y_\nu \overline{\psi_{L}}\; \phi \; \psi_{R}  +
\tilde{Y}_\nu \overline{\psi_{L}}\; \tilde{\phi}\; \psi_{R}  + Y_M
(\overline{\psi_{L}^c}\;  i \tau_2 \Delta_{L} \psi_L+
\overline{\psi_{R}^c}\;  i \tau_2 \Delta_{R} \psi_R) + h.c. \; ,
\end{eqnarray}
where $\psi_{L,R} = (\nu_{L,R}, l_{L,R})^T$. After the spontaneous
symmetry breaking, one may obtain the effective (light and
left-handed) neutrino mass matrix $m_\nu$ via the type II seesaw
mechanism:
\begin{equation}
m_\nu \; = \sqrt{2} (Y_M v_L - \frac{Y_D^2 \kappa^2}{2 Y_M v_R} )
\;, \label{seesaw 1}
\end{equation}
where $\kappa = \sqrt{|\kappa_1|^2 + |\kappa_2|^2} \approx 246$ GeV
represents the electroweak symmetry breaking (EWSB) scale and $Y_D
=(Y_\nu \kappa_1 + \tilde{Y}_\nu \kappa_2)/(\sqrt{2} \kappa)$. The
charged lepton mass matrix is given by $m_l =(Y_\nu \kappa_2 +
\tilde{Y}_\nu \kappa_1)/\sqrt{2}$. The electroweak precision test
requires $v_L \ll \kappa$. Barring extreme fine-tuning, the neutrino
masses $m_\nu \sim 0.1$ eV \cite{PDG} forces $v_L$ to be of order a
few eV or less, thereby requiring $v_R \sim 10^{12}$ GeV for $Y_D
\sim Y_M \sim m_l/\kappa$. In this case, the right-handed gauge
bosons $Z_2$ and $W_2$ are too heavy to be detected at the LHC and
the future colliders.  To allow for the possibility of an observable
right-handed scale, many authors focus on the $v_R \sim 10 $ TeV
case. Although the seesaw mechanism can work well, we need to
resolve the so-called VEV-seesaw puzzle \cite{VEV}, which is
indicated by a simple vacuum minimization equation:
\begin{equation}
(2\rho_1-\rho_3)v_L v_R=\beta_1 \kappa_1 \kappa_2+\beta_2
\kappa_1^2+\beta_3 \kappa_2^2 \;. \label{vev see-saw}
\end{equation}
Without loss of generality,  one can write Eq.(\ref{vev see-saw}) in
a compact form:
\begin{equation}
\gamma \equiv \frac{\beta}{\rho} = \frac{v_L v_R}{\kappa^2} \;.
\end{equation}
In view of the naturalness, one expects $\gamma \sim {\cal O }(1)$.
However, we find that $\gamma \sim 10^{-10}$ as long as $v_R \sim 10
$ TeV. This is the infamous  VEV-seesaw problem in the literatures
\cite{VEV}. The neutrino mass matrix $m_\nu $ in Eq. (\ref{seesaw
1}) can also be written as
\begin{equation}
m_\nu \; = \sqrt{2} \left(Y_M \gamma - \frac{Y_D^2 }{2 Y_M } \right
) \frac{\kappa^2}{v_R} \;. \label{seesaw 2}
\end{equation}

It is shown that the VEV-seesaw relationship implies the
unnaturalness for the auxiliary parameter $\gamma$ if one wants to
search for new physics at TeV scale. To avoid the VEV-seesaw puzzle,
a smart way is to introduce some new symmetries to eliminate all
$\beta$-type terms of the Higgs potential. However this is not a
easy task in the current model. One may guess there exists some
additional global symmetries like $U(1)$ acting on the Higgs fields
which can eliminate all $\beta$-type terms \cite{VEV}. However, such
alternative always affects the fermion sector and fails to give
correct fermion masses and mixing. If there is an approximate $U(1)$
horizontal symmetry to suppress $\beta_i$ without eliminating them
completely, then one may solve the VEV-seesaw problem \cite{Perez,
Kie}. Unfortunately, this model yields a small mixing angle within
the first two lepton generations. In Ref.\cite{VEV}, the authors
suggest a ${\cal Z}_2$ symmetry
\begin{eqnarray}
\Delta_L\rightarrow -\Delta_L,\hspace{1cm}\Delta_R\rightarrow
\Delta_R \;,
\end{eqnarray}
which can eliminate all $\beta$-type terms of the Higgs potential.
However, this discrete symmetry also eliminates the Majorana Yukawa
couplings, which implies that neutrinos are Dirac particles. At this
moment, Eq.(\ref{vev see-saw}) becomes
\begin{equation}
(2\rho_1-\rho_3) v_L = 0 \;.
\end{equation}
One may immediately dismiss the possibility $2\rho_1-\rho_3=0$,
which implies two massless left-handed Higgs triplet bosons. Thus
the only left choice is  $v_L = 0$. The ${\cal Z}_2$ symmetry leads
to $v_L = 0$ and the absence of both $\beta$-type terms and Majorana
Yukawa couplings. Furthermore, we find that the lightest particles
among the members of left-handed Higgs triplet $\Delta_L$, namely
$\delta_L^0$ and ${\delta_L^0}^*$, are two degenerate and stable
particles. A natural idea is that $\delta_L^0$ and ${\delta_L^0}^*$
may be the cold dark matter candidates. In the following sections we
shall discuss the possibility of $\delta_L^0$ and ${\delta_L^0}^*$
being the cold dark matter candidates  by evaluating  all relevant
annihilation processes. The main features of the LR symmetric model
with ${\cal Z}_2$ symmetry have been shown in Ref.\cite{DUKA}. Here,
we show the mass spectrum for the Higgs bosons and gauged bosons at
leading order in Table. I, with approximations $\kappa^2/v_R^2\simeq
0$ and $\kappa_2/\kappa_1\simeq 0$ mentioned in Appendix A. Gauge
bosons $Z_1$ and $Z_2$ are defined by $Z_1=c_W W_{3L}-s_W t_W
W_{3R}-\sqrt{c_{2W}}t_W B$ and $Z_2=\sqrt{c_{2W}} \sec_W W_{3R}-t_W
B$, where the subscript $W$ denotes the Weinberg angle $\theta_W$.
In addition, all the trilinear and quartic scalar interactions and
scalar-gauge interactions are listed in Appendix A for convenience.

\begin{table}
\begin{center}
\begin{tabular}{|l|l||l|l|}
 \hline Particles  & Mass$^2$ & Particles  & Mass$^2$\\\hline
$h^0 = \phi_1^{0 r}$ &  $m_{h^0}^2 = 2 \lambda_1 \kappa^2$ & $H_1^{\pm} = \phi_1^{\pm}$ &  $m_{H_1^\pm}^2 = \frac{1}{2} \alpha_3 (v_R^2 +\frac{1}{2} \kappa^2 )$ \\
$H_1^0 = \phi_2^{0 r}$ &  $m_{H_1^0}^2 = \frac{1}{2} \alpha_3 v_R^2 + 2 \kappa^2 (2 \lambda_2 +\lambda_3)$ & $\delta_R^{\pm \pm}$ &   $m_{\delta_R^{\pm \pm}}^2 = 2 \rho_2 v_R^2 +  \frac{1}{2} \alpha_3 \kappa^2 $ \\
$A_1^0 = - \phi_2^{0 i}$ &  $m_{A_1^0}^2 = \frac{1}{2} \alpha_3 v_R^2 - 2 \kappa^2 (2 \lambda_2 -\lambda_3)$ & $\delta_L^{\pm}$ &   $m_{\delta_L^{\pm}}^2 = \frac{1}{2} (\rho_3 - 2 \rho_1) v_R^2 +  \frac{1}{4} \alpha_3 \kappa^2 $ \\
$H_2^0 = \delta_R^{0 r}$ &  $m_{H_2^0}^2 = 2 \rho_1 v_R^2 $ & $\delta_L^{\pm \pm}$ &   $m_{\delta_L^{\pm \pm}}^2 = \frac{1}{2} (\rho_3 - 2 \rho_1) v_R^2 +  \alpha_3 \kappa^2 $ \\
$\delta_L^{0}$, ${\delta_L^{0}}^*$ &   $m^2 = \frac{1}{2} (\rho_3 -
2 \rho_1) v_R^2 $ & & \\ \hline \hline  $Z_1$ & $m_{Z_1}^2 =
\frac{g^2 \kappa^2}{4 \cos^2 \theta_W}$
&$W_1^\pm = W_L^\pm$ & $m_{W_1}^2 = \frac{g^2 \kappa^2}{4}$ \\
$Z_2$ & $m_{Z_2}^2 = \frac{g^2 v_R^2 \cos^2 \theta_W}{\cos 2
\theta_W}$ &$W_2^\pm = W_R^\pm$ & $m_{W_2}^2 = \frac{g^2 v_R^2}{2}$
\\\hline
\end{tabular}
\end{center} \caption{The mass spectrum for the Higgs bosons and the gauged
bosons in the LR symmetric model with ${\cal Z}_2$ symmetry. Here,
we have neglected the terms in order of $\kappa_2/\kappa_1$ and
$\kappa^2/v_R^2$. }\label{Phy.def.}
\end{table}

\section{The direct dark matter detection}

The current direct dark matter detection experiments, such as the
CDMS\cite{CDMS} and XENON\cite{XENON}, have provided very strong
constraints on the WIMP-nucleus elastic cross section. The rate for
direct detection of dark matter candidates is given by \cite{DM}
\begin{eqnarray}
R \approx \sum_i N_i \; \frac{\rho_{local}}{m}  \; \langle \sigma_{i
\cal N} \rangle\; , \label{Direct}
\end{eqnarray}
where $N_i$ is the number of  nuclei with species $i$ in the
detector, $\rho_{local}$ is the local energy density of dark matter,
$m$ is the mass of cold dark matter. $\sigma_{i \cal N}$ is the
WIMP-nucleus elastic cross section, and the angular brackets denote
an average over the relative WIMP velocity with respect to the
detector. Using the standard assumptions of $\rho_{local}$ and
distribution of the relative WIMP velocity \cite{Halo}, one can
derive the constrains on WIMP-nucleon cross-section
$\sigma_{n}^{exp} \leq 4.6 \times 10^{-44} \, {\rm cm^2}$ for $m =
60 \, {\rm GeV}$ from the CDMS \cite{CDMS}; $\sigma_{n}^{exp} \leq
8.8 \times 10^{-44} \, {\rm cm^2}$ for $m = 100 \, {\rm GeV}$ from
the XENON \cite{XENON}. Since the WIMP flux decreases $\propto 1/m$,
$\sigma_{n}^{exp} \propto m$ is a very good assumption for $m > 100
\, {\rm GeV}$.

In our scenario, the dark matter candidates $\delta_L^0$ and
${\delta_L^0}^*$ interact with nucleus ${\cal N}$ through their
couplings with quarks by exchanging the neutral gauge bosons $Z_1$,
$Z_2$ and Higgs bosons. We find that the main contribution comes
from the $Z_1$ exchanging process, which produces a spin-independent
elastic cross section on a nucleus ${\cal N}$ \cite{MDM}
\begin{eqnarray}
\sigma_{\cal N} = \frac{2 G_F^2 M^2({\cal N})}{\pi} [(A-Z) - (1- 4
\sin^2 \theta_W)Z]^2 \; ,
\end{eqnarray}
where $Z$ and $A-Z$ are the numbers of protons and neutrons in the
nucleus, respectively. $G_F$ is Fermi coupling constant and $M({\cal
N}) = m M_{\cal N}/(m + M_{\cal N})$ is the reduced WIMP mass.
Traditionally, the results of WIMP-nucleus elastic experiments are
presented in the form of a normalized the WIMP-nucleon cross section
$\sigma_{n}$ in spin-independent case, which is straight forward
\begin{eqnarray}
\sigma_{n} =\frac{1}{A^2} \frac{M^2(n)}{ M^2({\cal N})} \sigma_{\cal
N} \; ,
\end{eqnarray}
where $M(n) = m M_{n}/(m + M_{n})$ and $M_n$ denotes the nucleon
mass. When $m \gg M_{n}$, one may arrive at $\sigma_{n} = 8.2 \times
10^{-39} {\rm cm^2}$ for the CDMS experiment, which is $3-5$ orders
of magnitude above the present bounds for $m \sim 10^2-10^4 $ GeV
\cite{XENON}. Therefore, such dark matter candidates are excluded by
the current direct detection experiments.

If $\delta_L^0$ and ${\delta_L^0}^*$ have a nonzero splitting, one
can avoid the above bounds since the $Z_1$  exchanging  process is
forbidden kinematically \cite{KIN}. However, such degeneracy can not
be satisfied in our model. If the energy density of $\delta_L^0$ and
${\delta_L^0}^*$ in the solar system is far less than
$\rho_{local}$, we can avoid the above experimental limits as shown
in Eq.(\ref{Direct}). This means that $\delta_L^0$ and
${\delta_L^0}^*$ are only a very small part of the total dark
matter. We find that our model is consistent with the direct
detection experiments only when
\begin{eqnarray}
n_{\delta_L^0} \leq 4.8 \times 10^{-14} \; , \label{n}
\end{eqnarray}
where $n_{\delta_L^0}$ is the total relic number density of
$\delta_L^0$ and ${\delta_L^0}^*$. Here we have taken the
approximation $\sigma_n^{exp} \propto m$ (when $m \geq 100$ GeV) and
used $\sigma_n^{exp} = 3.4 \times 10^{-43} {\rm cm^2}$ ( $m = 1 \,
{\rm TeV}$) as the input parameter \cite{CDMS}. It is worthwhile to
stress that the bound in Eq.(\ref{n}) is not valid for $m < 100$
GeV.

The present experimental bounds are based on the standard
assumptions for the galatic halo \cite{Halo}. It needs to be
mentioned that the rotation curves of disk galaxies may also be
explained by the Modified Newtonian Dynamics (MOND) \cite{MOND}. On
one hand, we use the MOND to account for the rotation curve of the
Milk Way; On the other hand, we still believe that the cold dark
matter exists in the universe. In this case, the local energy
density of cold dark matter may be far less than the standard
assumption. Therefore, we may give up the above constraints from the
direct dark matter detection experiments. Subsequently,  the stable
particles $\delta_L^0$ and ${\delta_L^0}^*$ may be the cold dark
matter.

\section{Constraints on the annihilation cross section}

The thermal average of  annihilation cross section times the
``relative velocity"  $\langle \sigma v \rangle$ is a key quantity
in the determination of the cosmic relic abundances of $\delta_L^0$
and ${\delta_L^0}^*$. The constraint in Eq.(\ref{n}) implies
$\langle \sigma v \rangle$ must be very large in our scenario. In
this section, we analyze whether the present model can satisfy
Eq.(\ref{n}).

In our scenario, $\delta_L^i$ ($i=1,... ,6$ for $\delta_L^0$,
${\delta_L^0}^*$, $\delta_L^\pm$ and $\delta_L^{\pm \pm}$) are a set
of similar particles whose masses may be nearly degenerate. The
total relic density of the lightest particles  $\delta_L^0$ and
${\delta_L^0}^*$ is determined not only by their annihilation cross
sections, but also by the annihilation of the heavier particles,
which will later decay into $\delta_L^0$ or ${\delta_L^0}^*$.
Therefore, we need to consider the coannihilation processes
\cite{CO}. Since $\delta_L^\pm$ and $\delta_L^{\pm \pm}$ which
survive annihilation eventually decay into $\delta_L^0$ or
${\delta_L^0}^*$, the relevant quantity is the total number density
of $\delta_L^i$, $n = \sum_{i =1}^6 n_i$. The evolution of $n$ is
given by the following Boltzmann equation \cite{CO}:
\begin{eqnarray}
\frac{d n}{d t} = - 3 H n - \langle \sigma_{eff} v \rangle (n^2
-n_{eq}^2) \; , \label{BOL}
\end{eqnarray}
where $H$ is the Hubble parameter, $n_{eq}$ is the total equilibrium
number density, $v$ is the relative velocity of two annihilation
particles. The effective annihilation cross section $\sigma_{eff}$
is
\begin{eqnarray}
\sigma_{eff} = \sum_{i j}^6 \sigma_{i j} \frac{g_i g_j}{g_{eff}^2}(1
+ \Delta_i)^{3/2} (1 + \Delta_j)^{3/2} e^{- x (\Delta_i+ \Delta_j)}
\; ,
\end{eqnarray}
where $\Delta_i = (m_i - m)/m$, $x \equiv m/T$ is the scaled inverse
temperature. $g_i =1$ is the internal degrees of freedom of
$\delta_L^i$ and $g_{eff} = \sum_{i=1}^6 g_i (1 + \Delta_i)^{3/2}
e^{- x \Delta_i}$. For the total equilibrium number density, we may
use the nonrelativistic approximation $n_{eq} \approx g_{eff} (m T/
2 \pi)^{3/2} {\rm exp}(-m/T)$.

For particles which potentially play the role of cold dark matter,
the relevant freeze-out temperature is $x_f = m/T \sim 25$. In our
scenario, one can derive $x_f \gtrsim 35$ which can be seen in
Eq.(\ref{xf}). When $\Delta_i > 0.1$ for $\delta_L^\pm$ and
$\delta_L^{\pm \pm}$, we can arrive at $\sigma_{eff} =
\sigma_{12}/2$ in our model. In addition, we find that it is also a
rational approximation $\sigma_{eff} \approx \sigma_{12}/2$ even if
all masses of $\delta_L^i$ are nearly degenerate. For simplicity, we
take $\langle \sigma_{eff} v \rangle = \langle \sigma_{12} v \rangle
/2$ in the remaining analysis of our paper.

For nonrelativistic gases, the thermally averaged annihilation cross
section $\langle \sigma_{12} v \rangle$ may be expanded in powers of
$x^{-1}$, $\langle \sigma_{12} v \rangle = \sigma_0 x^{-k}$, $k=0$
for the $s$-wave annihilation and $k=1$ for the $p$-wave
annihilation \cite{KOLB}. The general formula for  $\langle
\sigma_{12} v \rangle$ is given by \cite{APP}
\begin{eqnarray}
\langle \sigma_{12} v \rangle = \sigma_0 x^{-k} = \frac{1}{m^2}
\left [  \omega - \frac{3}{2} (2 \omega - \omega ')x^{-1} + \ldots
\right]_{s/4m^2=1} \; , \label{Expand}
\end{eqnarray}
where $\omega \equiv E_1 E_2 \sigma_{12} v$, prime denotes
derivative with respect to $s/4m^2$, and $s$ is the center-of-mass
squared energy. $\omega$ and its derivative are all to be evaluated
at $s/4m^2=1$. The final number density $n_{\delta_L^0}$ is given by
\cite{KOLB}
\begin{eqnarray}
n_{\delta_L^0} = 2970 \, \frac{3.79 (k+1) x_f^{k+1}}{g_*^{1/2}
M_{Pl} \, m \, \sigma_0/2} \, {\rm cm}^{-3} \label{nin}
\end{eqnarray}
with
\begin{eqnarray}
x_f & = & {\rm ln}[0.038(k+1)(g_{eff}/g_*^{1/2}) M_{Pl}  \, m \,
\sigma_0/2] \nonumber \\ & - & (k+ 1/2)\, {\rm ln} \{ {\rm ln}
[0.038(k+1)(g_{eff}/g_*^{1/2}) M_{Pl} \, m \, \sigma_0/2] \} \; ,
\label{xf}
\end{eqnarray}
where $M_{Pl} = 1.22 \times 10^{19}$ GeV and $g_*$ is the total
number of effectively relativistic degrees of freedom at the time of
freeze-out. Here we take $g_* \approx 100$ for illustration. With
the help of Eqs. (\ref{n}), (\ref{nin}) and (\ref{xf}), we can
derive
\begin{eqnarray}
m \, \sigma_0  & \geq &  0.13 \; {\rm GeV}^{-1} \;\;\; (s{\rm -wave}) \; ; \nonumber \\
m \, \sigma_0  & \geq  & 9.8  \; {\rm GeV}^{-1} \;\;\; (p{\rm
-wave}) \;. \label{msigma}
\end{eqnarray}

\section{One Higgs bi-doublet model}

In this section, we shall investigate whether the above bounds can
be satisfied in one Higgs bi-doublet model or not. Since there are
many unknown parameters, some rational assumptions have to be made
for our model so that one can calculate all relevant annihilation
processes. In our scenario, the thermally averaged annihilation
cross section $\langle \sigma_{12} v \rangle$ is usually inverse
proportional to $m^2$ as shown in Eq. (\ref{Expand}). Therefore, one
can obtain $m \sigma_0 \propto 1/v_R$. Namely, the smaller $v_R$ is,
the easier Eq.(\ref{msigma}) can be satisfied. Considering the
constraints on the masses of $W_2$ and the FCNC Higgs boson from low
energy phenomenology \cite{Ji}, we  choose $v_R = 10$ TeV and
$\alpha_3 = 2$ as an instructive example to illustrate the main
features of our scenario. One can immediately get $m_{Z_2} = 7.5$
TeV and $m_{W_2} = 4.5$ TeV. Now let's introduce an auxiliary
parameter $\varepsilon \equiv (\rho_3 - 2 \rho_1)/(2 \rho_1)$ to
reexpress the mass of $\delta_L^0$ and ${\delta_L^0}^*$
\begin{eqnarray}
m = \sqrt{\frac{1}{2} (\rho_3 -2 \rho_1)} \, v_R = \sqrt{\varepsilon
\rho_1} v_R \; .
\end{eqnarray}
From the $Z_1$ invisible width one may obtain $m > m_{Z_1}/2$, which
requires $\varepsilon \rho_1 > 2.0 \times 10^{-5}$. On the other
hand,  we may require  $\rho_3 \leq 4$ in view of the
perturbativity, and then  derive $\rho_1 + \varepsilon \rho_1 \leq
2$. In addition,  we wish all $\rho_i$ have the same order which
means $ \varepsilon \leq 4$. Due to the suppression of phase space,
one may ignore some annihilation processes in terms of the values of
$\varepsilon$ and $\rho_1$. When $m < m_{W_1}$,  $\delta_L^0$ and
${\delta_L^0}^*$ mainly annihilate into the fermion pairs (except
for top quark). The corresponding $m \sigma_0$ is far less than the
lower bound of Eq.(\ref{msigma}). For the convenience of the
remaining analysis, we require $m \geq 500$ GeV (Namely,
$\varepsilon \rho_1 \geq 2.0 \times 10^{-3}$) which does not affect
our conclusions. Finally, we assume that all $\alpha_i$ of the Higgs
potential have the same order.

It is worthwhile to stress that  Eq.(\ref{Expand}) is not valid when
the annihilation takes place near a pole in the cross section
\cite{CO}. This happens, for example, in $Z$-exchange annihilation
when the mass of relic particle is near $m_{Z}/2$. For the cases
$2m/m_{Z} \leq 0.8$ and $2m/m_{Z} \geq 1.2$, we use the above
analytic way to calculate $m \sigma_0$. On the contrary, we should
numerically solve the Boltzmann equation in Eq.(\ref{BOL}), in which
the resonant cross sections of the Breit-Wigner form must be
considered. Then one can derive the relic number density
$n_{\delta_L^0}$ which has to be less than the upper bound in
Eq.(\ref{n}).

In general, all relevant annihilation processes may be divided into
four categories in terms of the different final states: $\delta_L^0
{\delta_L^0}^* \rightarrow f \bar{f}$, $\delta_L^0 {\delta_L^0}^*
\rightarrow V V$, $\delta_L^0 {\delta_L^0}^* \rightarrow H H$ and
$\delta_L^0 {\delta_L^0}^* \rightarrow V H$, where $V$ and $H$
denote the gauge boson and the Higgs boson, respectively. Next, we
shall analyze in detail the four classes of annihilation processes
and the resonance case.

\subsection{$\delta_L^0 {\delta_L^0}^* \rightarrow f \bar{f}$}

Let's start with the first case:  $\delta_L^0$ and ${\delta_L^0}^*$
annihilate into fermion pairs. There are two kinds of Feynman
diagrams at the tree level contributing to this case: S channel
gauge bosons exchanging and Higgs bosons exchanging diagrams.
Because of the absence of Majorana-type Yukawa couplings, there are
no the T channel diagrams' contribution. The first amplitude is
proportional to $e^2$, while the second is proportional to $\alpha
m_f/\sqrt{s}$. It is plausible that both diagrams have the same
contribution for $m\sim 500$ GeV. However, the squared amplitude of
the first diagram always includes a suppression factor of $1 - 4
m^2/s$, which leads to the $p$-wave annihilation.

For the gauge bosons exchanging diagram, we can obtain
\begin{eqnarray}
\omega_{f\bar{f}} \approx E_1 E_2 \sigma_{f\bar{f}} v  \approx
\frac{e^4}{4 \pi} \left ( 1- \frac{4 m^2}{s}\right ) \frac{9 s^2 -
19 m_{Z_2}^2 s + 11 m_{Z_2}^4}{(s - m_{Z_2}^2)^2}\;.
\end{eqnarray}
It is obvious that this  is a $p$-wave annihilation process. With
the help of Eq.(\ref{Expand}), we have $(m \sigma_0)_{f\bar{f}} \leq
2.2 \times 10^{-5} \; {\rm GeV}^{-1}$ for $m \geq 500$ GeV. It is 6
orders less than the lower bound $m  \sigma_0   \geq  9.8 \; {\rm
GeV}^{-1}$. Although one may increase $m \sigma_0$ through lowering
$m$,  $(m \sigma_0)_{f\bar{f}}$ is still far less than the lower
bound in Eq.(\ref{msigma}) even if $m = 100$ GeV. Therefore, this
process can not suppress the relic number density of $\delta_L^0 $
and ${\delta_L^0}^*$.

For the Higgs bosons exchanging diagram, the exchanged particles
should be $h^0$ and $H_1^0$. As shown in Table I, the mass of
$H_1^0$ is far more than the light SM Higgs mass $m_{h^0}$. Due to
the suppression of propagator, we neglect the contribution from
$H_1^0$. For the  $h^0$ case, the amplitude of Higgs bosons
exchanging process is proportional to $m_f$. Furthermore, we only
consider the top quark pair final states. The relevant cross section
is
\begin{eqnarray}
\omega_{top} \approx \frac{3}{16 \pi} \frac{(\alpha_1 m_t)^2 s}{(s -
m_{h^0}^2)^2} \;,
\end{eqnarray}
which leads to a $s$-wave annihilation process. One may immediately
derive $(m \sigma_0)_{top} \leq 1.5 \times 10^{-5} \; {\rm
GeV}^{-1}$ for $m \geq 500$ GeV and $\alpha_1 =2$, which is far less
than the lower bound $m \sigma_0 \geq 0.13 \; {\rm GeV}^{-1}$.

\subsection{ $\delta_L^0 {\delta_L^0}^* \rightarrow V V$}

\begin{figure}[h]\begin{center}
\includegraphics[scale=0.5]{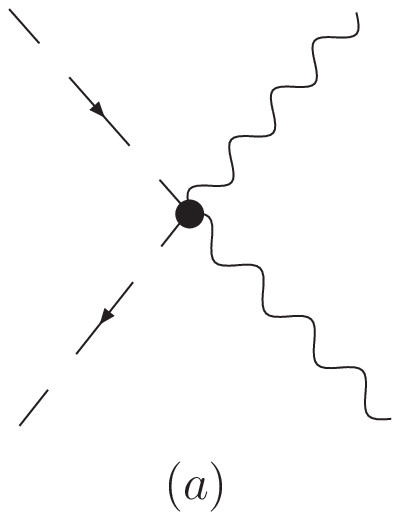}\hspace{0.8cm}
\includegraphics[scale=0.5]{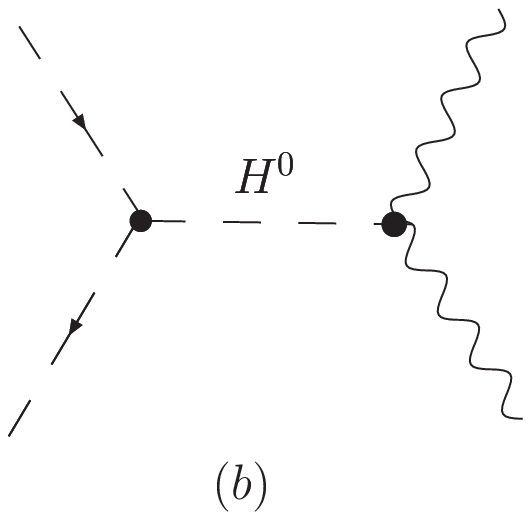}\hspace{0.8cm}
\includegraphics[scale=0.5]{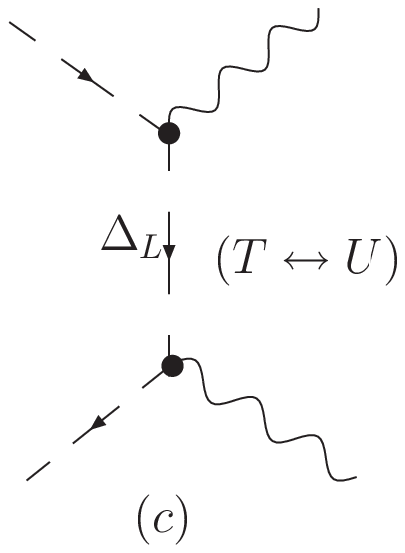}\hspace{0.8cm}
\includegraphics[scale=0.55]{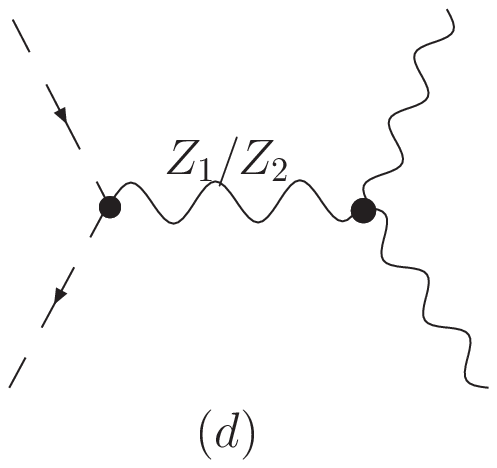}
\end{center}
\caption{All possible Feynman diagrams for the annihilation
processes $\delta_L^0 {\delta_L^0}^* \rightarrow V V$, where
$\Delta_L$  may be $\delta_L^{0/0*}$ or $\delta_L^{\pm}$, and  $H^0$
denotes $h^0$, $H_1^0$ and $H_2^0$. } \label{FDVV}\end{figure}

In Fig. \ref{FDVV}, we show all possible Feynman diagrams for the
process $\delta_L^0 {\delta_L^0}^* \rightarrow  V V$. There are
three kinds of Feynman diagrams Fig. 1a, 1b and 1c for the final
states $Z_1 Z_1$. Obviously, the amplitude of Fig. 1b is suppressed
by a factor of $\kappa/\sqrt{s}$ compared with the first one. Thus
we only consider the contribution from Fig. 1a and 1c. The total
annihilation cross section is found to be
\begin{eqnarray}
\omega_{Z_1 Z_1} \approx \frac{2 e^4 \csc^4 2 \theta_W}{\pi} \left[1
+ \frac{4 m^2}{s} - \frac{8 m^2 (s - 2 m^2)}{s^2} y(x_1) \right]
\end{eqnarray}
where the function $y(x_1)$ is defined by $y(x_1) \equiv {\rm
arctanh}(x_1)/ x_1$ and $x_1 = \sqrt{1-4 m^2/s}$. Then, we can
derive $(m \sigma_0)_{Z_1 Z_1} \leq 2.1 \times 10^{-5}  \; {\rm
GeV}^{-1}$ for $m \geq 500$ GeV. It is obvious that this result is
not so large as to satisfy the requirement of Eq.(\ref{msigma}).

According to the $Z_1 Z_1$ experience, we also calculate the other
processes. The corresponding cross sections are given by
\begin{eqnarray}
\omega_{W_1 W_1} & \approx & \frac{e^4}{32 \pi \sin^4 \theta_W }
\left[3 + \frac{28 m^2}{s}-
32 \frac{m^2}{s} y(x_1) \right]\; ;  \\
\omega_{W_2 W_2} & \approx & \frac{e^4}{128 \pi \sin^4 \theta_W
}\left(1-\frac{4m^2_{W_2}}{s}\right)^{1/2}
\bigg[\frac{4}{3}\frac{\sin^4\theta_W}{\cos^22\theta_W}\frac{(s-4m^2)(s-4m^2_{W_2})}{(s-m^2_{Z_2})^2}
\nonumber\\
&\times &\left(1+\frac{20m^2_{W_2}}{s}+12\frac{m^4_{W_2}}{s^2}\right) +\left(\frac{\rho_3 v_R^2}{m^2_{W_2}}\right)^2\frac{s^2-4m^2_{W_2}s+12m^4_{W_2}}{(s-m^2_{H_2^0})^2}\bigg];\\
\omega_{Z_1 Z_2} & \approx & \frac{e^4 \sec^4 \theta_W
}{4 \pi \cos 2 \theta_W} \left(1 - \frac{ m^2_{Z_2} }{s}\right) \nonumber\\
& \times & \frac{s^2 -3 m^2_{Z_2}s+ m^4_{Z_2}+
4m^2s-2(s-2m^2)(4m^2-m^2_{Z_2})y(x_1)}{(s-m^2_{Z_2})^2}\;; \\
\omega_{Z_2 Z_2}& \approx & \frac{e^4 \tan^4\theta_W}{32 \pi \cos^2
2\theta_W}\left(1 - \frac{4 m^2_{Z_2}}{s}\right)^{1/2} \nonumber\\
&\times & \bigg[ 4 + \frac{(4m^2-m^2_{Z_2})^2}{m^2s-4m^2
m^2_{Z_2}+m^4_{Z_2}}-4x_{\rho}[2+\frac{8m^2-s-2m^2_{Z_2}}{s-2m^2_{Z_2}}y(x_2)]
\nonumber\\
&-&\frac{16(2m^2-m^2_{Z_2})s-4m^2(16m^2-7m^2_{Z_2})}{(s-2m^2_{Z_2})^2}
y(x_2)
+(6-\frac{2s}{m^2_{Z_2}}+\frac{s^2}{2m^4_{Z_2}})x^2_{\rho}\bigg],
\end{eqnarray}
where $x_{\rho}=(\cot^4\theta_w\rho_3v_R^2)/(s-m^2_{H_2^0})$ and
$x_2 \equiv \sqrt{(s-4m^2)(s-4m_{Z_2}^2)}/(s-2m_{Z_2}^2)$. These
cross sections have the same order as the $Z_1 Z_1$ case. However,
the thermally averaged annihilation cross sections of these
processes (except for $W_1 W_1$) are far less than the $Z_1 Z_1$
case with $m = 500$ GeV. Therefore, we don't analyze these processes
in detail.

\subsection{$\delta_L^0 {\delta_L^0}^* \rightarrow H H / V H$}

\begin{figure}[h]\begin{center}
\includegraphics[scale=0.45]{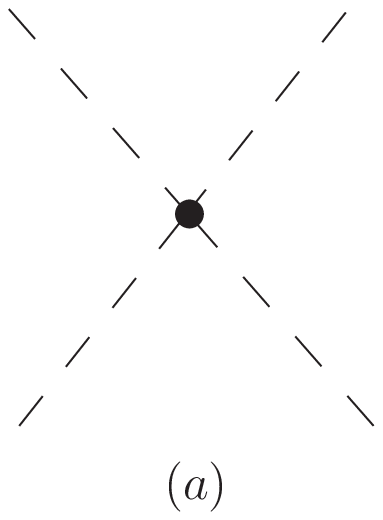}\hspace{0.8cm}
\includegraphics[scale=0.45]{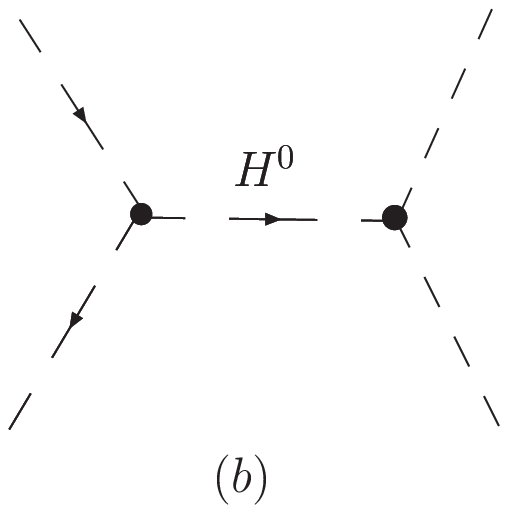}\hspace{0.8cm}
\includegraphics[scale=0.5]{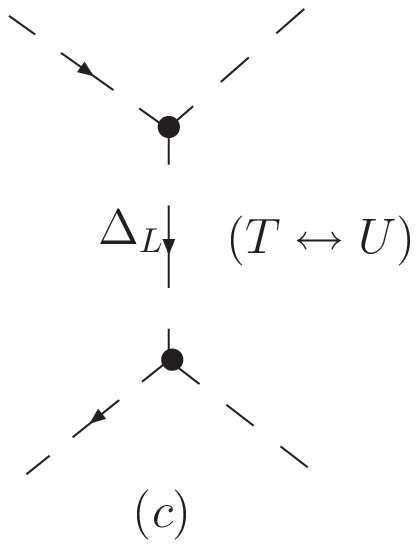}\hspace{0.8cm}
\includegraphics[scale=0.5]{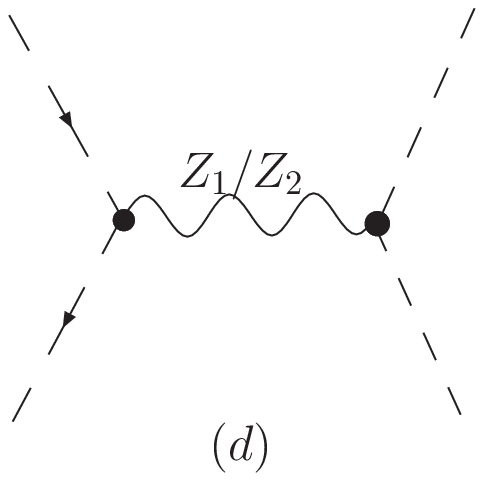}
\end{center}
\caption{All possible Feynman diagrams for the annihilation
processes $\delta_L^0 {\delta_L^0}^* \rightarrow H H$.}
\label{FDHH}\end{figure}

\begin{figure}[h]\begin{center}
\includegraphics[scale=0.45]{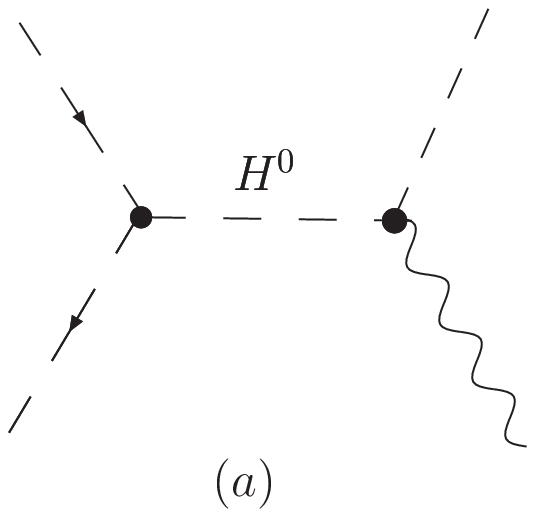}\hspace{0.8cm}
\includegraphics[scale=0.5]{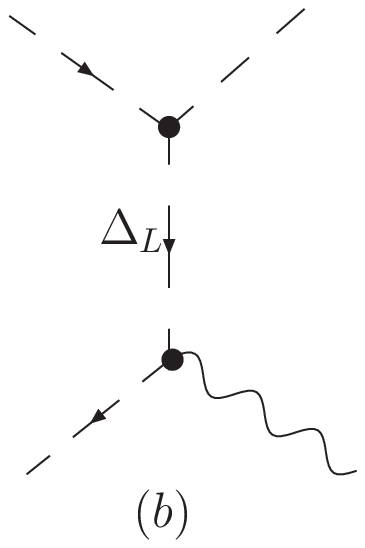}\hspace{0.8cm}
\includegraphics[scale=0.5]{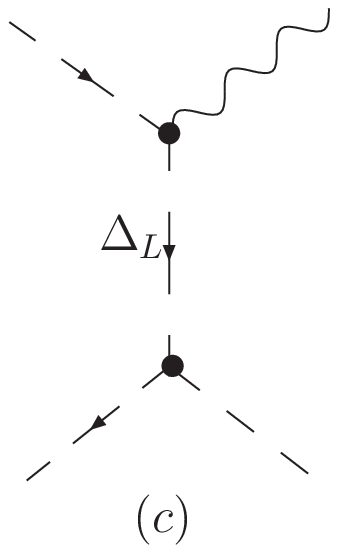}\hspace{0.8cm}
\includegraphics[scale=0.5]{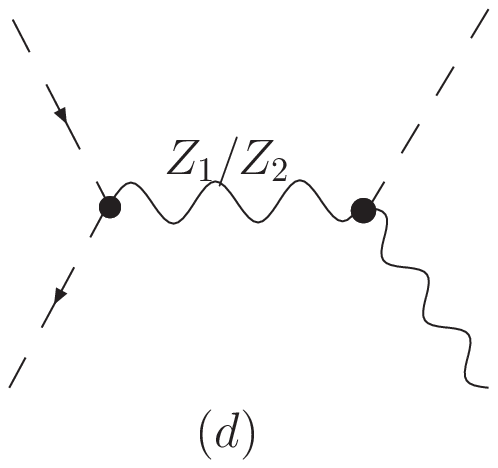}
\end{center}
\caption{All possible Feynman diagrams are shown for the
annihilation processes $\delta_L^0 {\delta_L^0}^* \rightarrow V H$.
The first diagram only appears in the process
$\delta_L^0\delta_L^{0*}\rightarrow V A_1^0$}
\label{FDHV}\end{figure}

Let's now focus on the processes $\delta_L^0 {\delta_L^0}^*
\rightarrow H H / V H$. The relevant Feynman diagrams for $HH$ and
$HV$ are shown in Fig. \ref{FDHH} and Fig. \ref{FDHV}, respectively.
Since the dimensional scalar trilinear couplings enter extensively
into the above two annihilation processes, the electroweak scale
coupling $\alpha_1 \kappa$ in $\delta_L^0 {\delta_L^0}^* h^0$ and
the right-handed scale coupling $\rho_3 v_R$ in $\delta_L^0
{\delta_L^0}^* H_2^0$ would make big difference in the $\delta_L^0
{\delta_L^0}^*\rightarrow H V$ processes according to our current
parameter setting. Considering the complexity of this model, we only
calculate the annihilation cross sections up to leading order (LO)
by omitting the next to leading order (NLO) contributions in terms
of the following three suppressing factors: (1) small VEV ratio
$\kappa/v_R$ and $\kappa/\sqrt{s}$ due to the big hierarchy in
symmetry breaking scale of the LR model. Since we have made the
approximation $\kappa^2/v_R^2\simeq 0$ thus here it is of course a
reasonable power counting rule to pick out the LO processes against
the NLO ones; (2) gauge coupling suppression $e^2$; (3) $p$-wave
factor $1-4m^2/s$ due to large suppression in the integration of
initial energy of dark matter pair.

\begin{table}
\begin{center}
\begin{tabular}{|c|l|c||c|l|c|}
 \hline Process & Amplitude & Order & Process & Amplitude & Order\\ \hline
 $h^0h^0$   & $i\alpha_1\left(1-\frac{\rho_3v_R^2}{s-2\rho_1v_R^2}\right)$  & 1 &$h^0Z_1$ &$4ie\alpha_1\csc2\theta_W\kappa P_{TU}$ &$\frac{e^2}{x_f}\frac{\kappa^2}{s}$\\
 $h^0H_1^0$ & $2i\alpha_2\left(1-\frac{\rho_3v_R^2}{s-2\rho_1v_R^2}\right)$ & 1 &$h^0Z_2^0$ & $2ie\alpha_1\frac{t_W}{\sqrt{c_{2W}}}\kappa P_{TU}$ & $\frac{e^2}{x_f}\frac{\kappa^2}{s}$ \\
 $H_1^0H_1^0$ & $i(\alpha_1+\alpha_3)\left(1-\frac{\rho_3v_R^2}{s-2\rho_1v_R^2}\right)$ & 1 & $H_1^0Z_1$ & $8ie\alpha_2\csc _{2W}\kappa P_{TU}$  & $\frac{e^2}{x_f}\frac{\kappa^2}{s}$\\
 $A_1^0A_1^0$ & $i(\alpha_1+\alpha_3)\left(1-\frac{\rho_3v_R^2}{s-2\rho_1v_R^2}\right)$ & 1 & $H_1^0Z_2$ & $4ie\alpha_2\frac{t_W}{\sqrt{c_{2W}}}\kappa P_{TU}$ & $\frac{e^2}{x_f}\frac{\kappa^2}{s}$\\
 $A_1^0H_1^0$ & $et_W\left[c_{2W}P_{Z_1}-t_WP_{Z_2}/2\right]$ & $\frac{e^2}{x_f}$ &$A_1^0Z_1$ & $4ie\alpha_2\csc _{2W}\kappa P^H_{43}$ & $1$\\
 $H_2^0H_2^0$ & $i\rho_3\left(2-\frac{6\rho_1v_R^2}{s-2\rho_1v_R^2}-\frac{\rho_3v_R^2}{t-m^2}-\frac{\rho_3v_R^2}{u-m^2}\right)$ & 1 & $A_1^0Z_2$ & $-4ie\alpha_2\csc _{2W}\sqrt{c_{2W}}\kappa P^H_{43}$ & $(\frac{\kappa}{v_R})^2$\\
 $h^0H_2^0$ & $-i\rho_3\alpha_1\kappa v_R\left(\frac{1}{s-2\rho_1v_R^2}+\frac{1}{t-m^2}+\frac{1}{u-m^2}\right)$ & $(\frac{\kappa}{v_R})^2$ & $H_2^0Z_1$ &$4ie\rho_3\csc _{2W}v_R P_{TU}$  & $\frac{e^2}{x_f}$\\
 $H_1^0H_2^0$ & $-2i\rho_3\alpha_2\kappa v_R\left(\frac{1}{s-2\rho_1v_R^2}+\frac{1}{t-m^2}+\frac{1}{u-m^2}\right)$ & $(\frac{\kappa}{v_R})^2$ & $H_2^0Z_2$ & $2ie\rho_3\frac{t_W}{\sqrt{c_{2W}}}v_R P_{TU}$ & $\frac{e^2}{x_f}$\\
 $H_1^+H_1^-$ & $i\alpha_1\left[1-(1+\frac{\alpha_3}{\alpha_1})\frac{\rho_3v_R^2}{s-2\rho_1v_R^2}\right]$ & 1& $H_1^+W_1^-$ & $-2ie \alpha_2\csc _{2W}\kappa P^H_{43}$ & 1\\
 $\delta_R^{++}\delta_R^{--}$ & $i\rho_3\left[1-\frac{2(\rho_1+2\rho_2)v_R^2}{s-2\rho_1v_R^2}-\frac{\rho_4}{\rho_3}\frac{8\rho_4v_R^2}{t-m^2}\right]$ & 1& $H_1^+W_2^-$ & $-2ie\alpha_2\csc _{2W}\kappa P^h_{43}$ & $(\frac{\kappa}{v_R})^2$\\ \hline
\end{tabular}
\end{center} \caption{The amplitude for $HH/HV$ final states,
where $c_W=\cos \theta_W$, $t_W=\tan \theta_W$, etc. We also
estimate the order of corresponding annihilation cross sections.
}\label{AMP}
\end{table}

In this subsection,  we apply the above three suppressing factors to
make an explicit demonstration of the LO processes, then give the
convincing dark matter annihilation cross sections. The LO amplitude
for each possible annihilation process is listed in Table \ref{AMP}.
The notations are as follows: $p_{1,2}$ denotes the momentum of the
dark matter pair, while $p_{3,4}$ is the momentum of the final
states, $\epsilon$ is the polar vector of gauge boson:
\begin{eqnarray}
P_{Z_{1,2}}&=&\frac{(p_1-p_2)\cdot(p_4-p_3)}{s-m^2_{Z_{1,2}}};
\hspace{0.4cm}P^{h,H}_{43}=\frac{p_4\cdot
\epsilon_(p_3)}{s-m^2_{h^0,H_1^0}};
\hspace{0.4cm}P_{TU}=\frac{p_2\cdot
\epsilon(p_3)}{t-m^2}-\frac{p_1\cdot \epsilon(p_3)}{u-m^2}.
\end{eqnarray}
Here we only consider the cross sections with amplitude order 1. In
terms of Table \ref{AMP},  nine LO annihilation  cross sections are
listed in Table \ref{cs}, where
$A=1-\frac{2(\rho_1+2\rho_2)}{s-2\rho_1v_R^2}$ and
$B=\frac{32\rho_4^2}{\rho_3} \frac{v_R^2}{s-2
m_{\delta_R^{\pm\pm}}^2}y(x_3)+ \frac{64\rho_4^4}{\rho_3^2}
\frac{v_R^4}{s m^2-4m^2 m^2_{\delta_R^{\pm\pm}}+
m^4_{\delta_R^{\pm\pm}}}$ with $x_3=\sqrt{(s-4
m^2)(s-4m^2_{\delta_{R}^{\pm\pm}})}
/(s-2m^2_{\delta_{R}^{\pm\pm}})$;
$a=2-\frac{3m^2_{H_2^0}}{s-m^2_{H_2^0}},b=4m^2+m^2_{H_2^0},
c=s-2m^2_{H_2^0},d=\sqrt{sm^2-4m^2m^2_{H_2^0}+m^4_{H_2^0}}$ and $
x_4=\sqrt{(s-4 m^2)(s-4m^2_{H_2^0})} /(s-2m^2_{H_2^0})$. We find
that these processes fail to provide enough large cross sections.
For $\delta_L^0 {\delta_L^0}^* \rightarrow A_1^0 Z_1$ and $H_1^\pm
W_1^\mp$,  one can easily obtain $m \sigma_0 \lesssim \alpha_2^2/(16
\pi m)$, which is far less than the required lower bound $0.13 \;
{\rm GeV}^{-1}$. Since the other processes have the similar forms,
we take the process $\delta_L^0 {\delta_L^0}^* \rightarrow h^0
H_1^0$ as an example to illustrate the main features of this kind of
processes. One can immediately derive
\begin{eqnarray}
(m \sigma_0)_{h^0 H_1^0} = \frac{\alpha_2^2}{8 \pi v_R}
\sqrt{\frac{1}{\varepsilon \rho_1}- \frac{1}{4(\varepsilon
\rho_1)^2}} \left( \frac{\varepsilon -2}{2 \varepsilon -1} \right)^2
\;,
\end{eqnarray}
where we have used $\alpha_3 =2$. It is obvious that the maximum
value can be obtained when $\varepsilon \rho_1=0.5$. Varying
$\varepsilon$, we may derive $(m \sigma_0)_{h^0 H_1^0} \leq
3.5\times 10^{-4} \; {\rm GeV}^{-1}$ for $\alpha_2 = 2$. Therefore,
we don't discuss this class of processes in detail.

\begin{table}
\begin{center}
\begin{tabular}{|c|l||c|l|}
\hline Process & $4 E_1E_2\sigma v $ & Process & $4 E_1E_2 \sigma v$
\\ \hline
$h^0h^0$ & $\frac{\alpha_1^2}{16\pi}(1-\frac{\rho_3v_R^2}{s-2\rho_1v_R^2})^2$ & $H_1^+H_1^-$ & $\frac{\alpha_1^2}{8\pi}\left[1-(1+\frac{\alpha_3}{\alpha_1})\frac{\rho_3v_R^2}{s-2\rho_1v_R^2}\right]^2$ \\
$h^0H_1^0$ & $\frac{\alpha_2^2}{2\pi}(1-\frac{m^2_{H_1^0}}{s})^{1/2}(1-\frac{\rho_3v_R^2}{s-2\rho_1v_R^2})^2$ & $\delta_R^{++}\delta_R^{--}$ & $\frac{\rho_3^2}{8\pi}(1-\frac{4m^2_{\delta_R^{\pm\pm}}}{s})^{1/2}(A^2+B)$ \\
$H_1^0H_1^0$ & $\frac{(\alpha_1+\alpha_3)^2}{16\pi}(1-\frac{4m^2_{H_1^0}}{s})^{1/2}(1-\frac{\rho_3v_R^2}{s-2\rho_1v_R^2})^2$ & $A_1^0Z_1$ & $\frac{\alpha_2^2}{4\pi}(1-\frac{m^2_{A_1^0}}{s})^{5/2}$ \\
$A_1^0A_1^0$ & $\frac{(\alpha_1+\alpha_3)^2}{16\pi}(1-\frac{4m^2_{H_1^0}}{s})^{1/2}(1-\frac{\rho_3v_R^2}{s-2\rho_1v_R^2})^2$ & $H_1^+W_1^-$ & $\frac{\alpha_2^2}{4\pi}(1-\frac{m^2_{H_1^{\pm}}}{s})^{5/2}$\\
$H_2^0H_2^0$ & $ \frac{\rho_3^2}{16 \pi} \left [ a^2+\frac{b^2}{2
d^2}+\frac{2b}{c}(2a+\frac{b}{c})y(x_4) \right ]$ & &
\\\hline
\end{tabular}
\end{center} \caption{The annihilation cross sections for the leading order processes. }\label{cs}
\end{table}

\subsection{The resonance case}

As pointed out in the previous discussion, the method of calculating
the effective thermally averaged annihilation cross section $\langle
\sigma_{eff} v \rangle$ is not valid for the resonance case
\cite{CO}. Here we numerically solve the Boltzmann equation
Eq.(\ref{BOL}), which can be reexpressed as \cite{Gondolo}
\begin{eqnarray}
\frac{dY}{dx} & = & -\frac{x}{H_{x=1} {\bf s}(x)} \gamma_{eff}
\left( \frac{Y^2}{Y^2_{\rm eq}}-1\right) \ ,
\end{eqnarray}
where $H_{x=1}$ is the Hubble parameter evaluated at $T = m$ and
${\bf s(x)}$ is the entropy density given by
\begin{eqnarray} H_{x=1} = \sqrt{\frac{4 \pi^3 g_{\ast}}{45}}\,
\frac{m^2}{M_{Pl}}\;, \quad {\bf s} (x)=\frac{2 \pi^2 g_{\ast}}{45}
\frac{m^3}{x^3}\; .
\end{eqnarray}
$Y \equiv n/{\bf s}$ is the ratio of the total particle number
density $n$ to the entropy density ${\bf s}$. The equilibrium number
density $Y_{\rm eq}$ reads
\begin{eqnarray}
Y_{\rm eq}(x)=\frac{45}{4\pi^4}\frac{g_{eff}}{g_{\ast}} x^2
K_2\left(x \right)\,.
\end{eqnarray}
In fact, $\gamma_{eff}$ is the reaction density defined by
\begin{eqnarray}
\gamma_{eff} \equiv n_{eq}^2 \langle \sigma_{eff} v \rangle =
\frac{m^4}{64\,\pi^4 x} \int_{4}^{\infty}
\hat{\sigma}_{eff}(z)\,\sqrt{z}\,K_1(x\,\sqrt{z})\,dz\,,
\end{eqnarray}
with
\begin{eqnarray}
\hat{\sigma}_{eff}= g_{eff}^2 4 E_1 E_2 \sigma_{eff} v \sqrt{1-4/z}
\;, \label{fac}
\end{eqnarray}
where $z =s /m^2$, $K_1(x)$ and $K_2(x)$ are the modified Bessel
functions.

\begin{table}
\begin{center}
\begin{tabular}{|l|c||l|c|}
 \hline $\rho_1 = 1.0$  &  $n_{\delta_L^0}$ & $\rho_1 = 0.1$  & $n_{\delta_L^0}$
 \\\hline
 $\alpha_1 = 0.01$ & $n_{\delta_L^0}=6.4 \times 10^{-13}$ &  $\alpha_1 = 0.01$ & $n_{\delta_L^0}=1.6 \times 10^{-12}$\\
 $\alpha_1 = 0.1$ & $n_{\delta_L^0}=6.5 \times 10^{-13}$ &  $\alpha_1 = 0.1$ & $n_{\delta_L^0}=1.4 \times 10^{-12}$\\
 $\alpha_1 = 1.0$ & $n_{\delta_L^0}=1.1 \times 10^{-12}$ &  $\alpha_1 = 1.0$ & $n_{\delta_L^0}=3.6 \times 10^{-12}$\\
 $\alpha_1 = 2.0$ & $n_{\delta_L^0}=2.5 \times 10^{-12}$ &  $\alpha_1 = 2.0$ & $n_{\delta_L^0}=1.2 \times 10^{-11}$\\\hline
\end{tabular}
\end{center}
\caption{The relic number density $n_{\delta_L^0}$ in terms of
different $\alpha_1$ and $\rho_1$ for the $H_2^0$ case.}
\label{Resonant}
\end{table}

In our scenario,  the exchanged particles may be $Z_1$, $Z_2$,
$h^0$, $H_1^0$ and $H_2^0$. It is obvious that the case of
exchanging gauge bosons $Z_1$ or $Z_2$ is a $p$-wave annihilation
process. If the exchanged particle is $H_1^0$, the corresponding
cross section will be suppressed by $\kappa^2/v_R^2$. For the $h^0$
case, the resonant condition $2 m \approx m_{h^0}$ implies that the
final states must be the Fermi pairs. In addition, the previous
analysis indicates that the maximal cross section might be from the
$H_2^0$ exchanging process. Therefore, we study the $h^0$ and
$H_2^0$ cases in this subsection.

Firstly we consider the $H_2^0$ case. Due to the factor
$\sqrt{1-4/z}$ in Eq. (\ref{fac}), we take $m^2_{H_2^0}/m^2 = 4.1$
(namely $\varepsilon = 0.4878$). At this point, $\gamma_{eff}$
becomes more larger than the $m^2_{H_2^0}/m^2 = 4$ case. Then we
take $\rho_1 = 1$ ($\rho_3 = 2.98$). At this moment, $\delta_L^0$
and ${\delta_L^0}^*$ may annihilate into $h^0 h^0$, $h^0 H_1^0$ and
$W_2 W_2$. Since $h^0 h^0$ and $h^0 H_1^0$ have the similar form,
the key quantity is $\alpha_2^2 + \alpha_1^2/8$ for our calculation.
Without loss of generality, one may take different values for
$\alpha_1$ and require $\alpha_1 = \alpha_2$. The final results for
different $\alpha_1$ have been shown in Table \ref{Resonant}. In
addition, we also calculate the $\rho_1 = 0.1$ case, and list the
corresponding results in Table \ref{Resonant}. If $\rho_1 = 0.1$,
the final states have to be two SM Higgs bosons. In views of Table
\ref{Resonant}, we may find that the $H_2^0$ case fails to suppress
the relic number density of $\delta_L^0 $ and ${\delta_L^0}^*$.

\begin{figure}[h]\begin{center}
\includegraphics[scale=0.45]{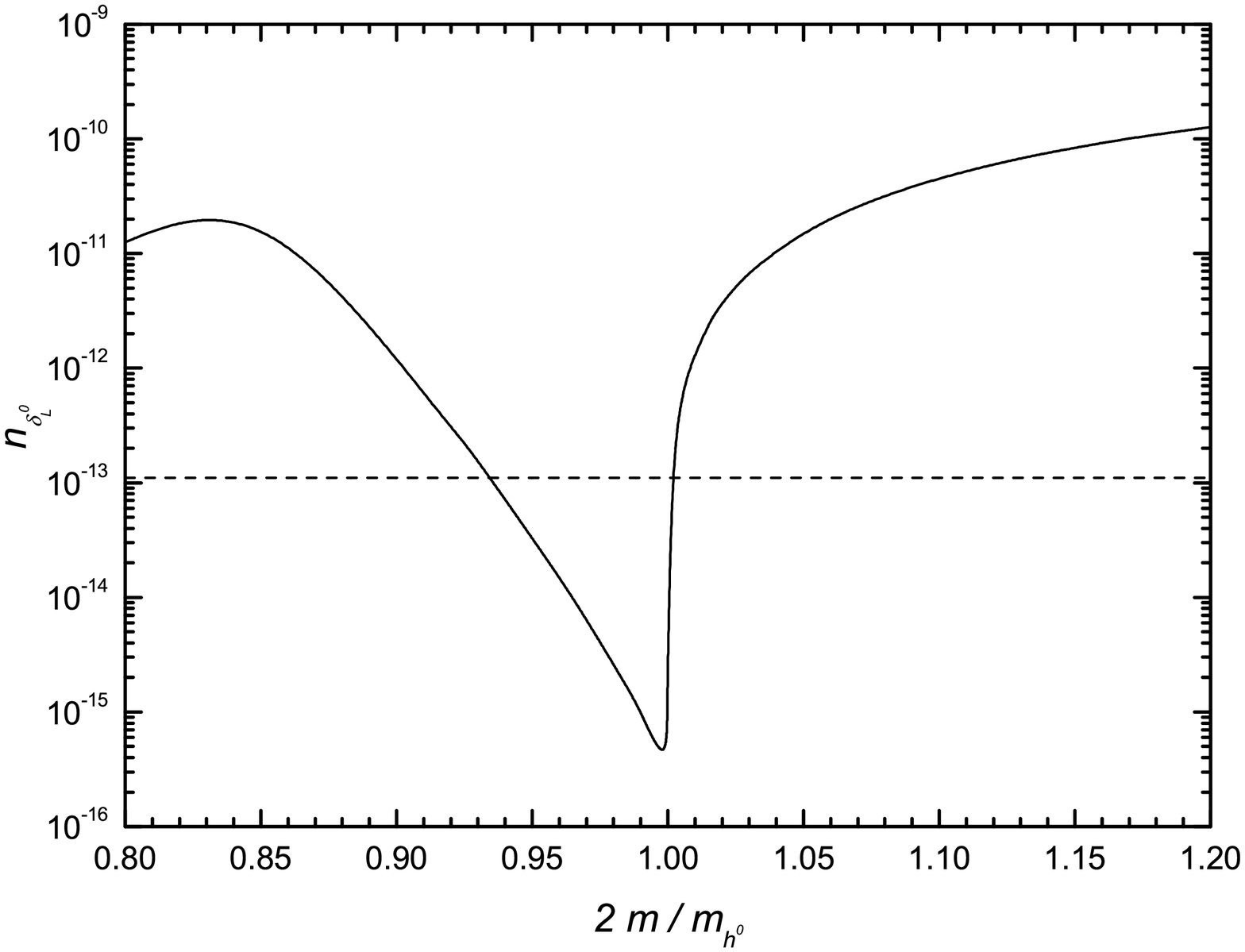}
\end{center}
\caption{Numerical illustration of the relic number density
$n_{\delta_L^0}$ as a function of $2 m / m_{h^0} $ near a resonance,
where $m_{h^0} = 120$ GeV has typically been taken. The dashed line
denotes the present experimental upper bound on $n_{\delta_L^0}$.}
\end{figure}

Now  we assume the SM Higgs mass $m_{h^0} = 120$ GeV in the $h^0$
case. Furthermore, one may obtain $m = 59.3$ GeV ($\varepsilon
\rho_1 = 3.5 \times 10^{-5}$) from $m^2_{H_2^0}/m^2 = 4.1$. Because
of $m < 100$ GeV, the bound in Eq.(\ref{n}) is not valid. For $m =
59.3$ GeV, we take $\sigma_{n}^{exp} \leq 4.6 \times 10^{-44} \,
{\rm cm^2}$ \cite{CDMS} and derive the corresponding bound
\begin{eqnarray}
n_{\delta_L^0} \leq 1.1 \times 10^{-13} \;.
\end{eqnarray}
In this case, $\delta_L^0 $ and ${\delta_L^0}^*$ mainly annihilate
into the bottom quark pair. The annihilation cross section is given
by
\begin{eqnarray}
(4 E_1 E_2 \sigma v)_{h^0} \simeq \frac{3}{4 \pi} \frac{\alpha_1^2
m_b^2 s}{(s-m_{h^0}^2)^2+m_{h^0}^2 \Gamma_{h^0}^2}
\end{eqnarray}
where $m_b$ is the bottom quark mass and $\Gamma_{h^0} \simeq 3
m_{h^0} m_b^2 /(8 \pi  \kappa^2)$ is the decay width of $h^0$. One
may obtain $n_{\delta_L^0} = 1.2 \times 10^{-15}$ for $\alpha_1 =2$.
This wonderful result indicates that our scenario may be consistent
with the direct dark matter search bound. To illustrate, we plot the
relic number density $n_{\delta_L^0}$ versus the dark matter mass
$m$ in Fig. 4, where all annihilation channels have been considered.
Using the results from CERN LEP-II, Datta and Raychaudhuri have
derived $m \geq 55.4$ GeV \cite{Datta}. To show the $h^0$ resonance
region, we choose 48 GeV $\leq m  \leq 72$ GeV ($0.8 \leq 2m/m_{h^0}
\leq 1.2$) in Fig. 4. The peak around $2m/m_{h^0} = 0.83$ in Fig. 4
is due to the competition between $h^0$ and $Z_1$ resonances. For
$m_{h^0} = 120$ GeV, we find that $56 {\rm GeV} \lesssim m \lesssim
60 {\rm GeV}$ can satisfy the requirement $n_{\delta_L^0} \leq 1.1
\times 10^{-13}$. At this moment, one may obtain
$\Omega_{\delta_L^0} h^2 \lesssim 6.3 \times 10^{-7}$, which is far
less than the total dark matter density $\Omega_{DM} h^2 = 0.111 \pm
0.006$ \cite{PDG}.

\section{Two Higgs bi-doublets model}

Motivated by the general two Higgs doublet model as a model for
spontaneous CP violation, one may simply extend the one Higgs
bi-doublet LR model to a two Higgs bi-doublets LR model with
spontaneous P and CP violation \cite{Wu}. Besides one left-handed
Higgs triplet $\Delta_L$ (3,1,2) and one right-handed  Higgs triplet
$\Delta_R$ (1,3,2), this model consists of two Higgs bi-doublets
$\phi$ (2,2,0) and $\chi$ (2,2,0), which can be written as
\begin{eqnarray}
\phi  = \left ( \matrix{ \phi_1^0 & \phi_1^+ \cr \phi_2^- & \phi_2^0
\cr  } \right ) ; \; \chi  = \left ( \matrix{ \chi_1^0 & \chi_1^+
\cr \chi_2^- & \chi_2^0 \cr  } \right ) \;.
\end{eqnarray}
The most general Yukawa interaction for quarks is given by
\begin{eqnarray}
-{\cal L}_Y  =   \overline{Q_L} \left ( Y_q \phi +\tilde{Y}_q
\tilde{\phi} + F_q \chi +\tilde{F}_q \tilde{\chi} \right) Q_R \; ,
\end{eqnarray}
where $Q_{L,R} = (u_{L,R}, d_{L,R})^T$. Parity P symmetry requires
$Y_q$, $\tilde{Y}_q$, $F_q$ and $\tilde{F}_q$ are hermitian
matrices. When both P and CP are required to be broken down
spontaneously, all the Yukawa couplings matrices are real symmetric.
After the spontaneous symmetry breaking, two Higgs bi-doublets can
have the following vacuum expectation values
\begin{eqnarray}
\langle \phi \rangle  = \left ( \matrix{ \kappa_1/\sqrt{2} & 0 \cr 0
& \kappa_2/\sqrt{2} \cr } \right ) ; \; \langle \chi \rangle  =
\left ( \matrix{ w_1/\sqrt{2} & 0 \cr 0 & w_2/\sqrt{2} \cr } \right
)\;,
\end{eqnarray}
where $\kappa_1$, $\kappa_2$, $w_1$ and  $w_2$ are in general
complex. Then we may obtain the following quark mass matrices
\begin{eqnarray}
M_u &=& \frac{1}{\sqrt{2}}(Y_q \kappa_1 + \tilde{Y}_q \kappa_2 + F_q
w_1 + \tilde{F}_q w_2) \;; \nonumber \\ M_d &=&
\frac{1}{\sqrt{2}}(Y_q \kappa_2 + \tilde{Y}_q \kappa_1 + F_q w_2 +
\tilde{F}_q w_1) \;.
\end{eqnarray}

In the two Higgs bi-doublets model, the stringent constraints from
the low energy phenomenology can be significantly relaxed. In Ref.
\cite{Wu}, the authors calculate the constraints from neural $K$
meson mass difference $\Delta m_{K}$ and demonstrate that a
right-handed gauge boson $W_2$ contribution in box-diagrams with
mass around 600 GeV is allowed due to a cancelation caused by a
light charged Higgs boson with a mass range $150 - 300$ GeV.
Therefore, we take $v_R \approx 2$ TeV instead of the previous $v_R
= 10$ TeV for this section. It is worthwhile to stress that our
previous estimation is still right for this case except for the
process $\delta_L^0 {\delta_L^0}^* \rightarrow h^0 H_1^0$. $(m
\sigma_0)_{h^0 H_1^0}$ in Eq.(30) will be about 5 times larger than
that in the $v_R = 10$ TeV case, which does not affect our
conclusion.

Since there are two Higgs bi-doublets, we can give more dark matter
annihilation processes for $\delta_L^0 {\delta_L^0}^* \rightarrow H
H$ and $\delta_L^0 {\delta_L^0}^* \rightarrow V H$. In this model,
one may obtain three light neutral Higgs bosons and a pair of light
charged Higgs bosons \cite{Wang}. The other Higgs bosons' masses are
related to $v_R$. Although the annihilation cross section might be
doubled or even increased by several times, it is still at least 10
times less than the direct dark matter search bound.

An significant advantage of the two Higgs bi-doublets model is that
the  Yukawa  couplings may become very large. In view of Eq. (41),
one can explicitly understand this feature. For example, we require
the couplings $Y_q$ and $\tilde{Y}_q$ are very large when $w_1 \gg
\kappa_1 \gg \kappa_2 \approx w_2$. Then one may obtain more larger
annihilation cross section for the $\delta_L^0 {\delta_L^0}^*
\rightarrow f \bar{f}$ process than Eq.(23). For illustration, we
take the maximal annihilation cross section for each quark pair
final states
\begin{eqnarray}
4 E_1 E_2 \sigma v \sim \frac{3}{8 \pi} \frac{(\alpha_i Y_q w_1)^2
s}{(s - m_{h}^2)^2} \;,
\end{eqnarray}
where $ m_{h}$ denotes the mass of a light Higgs boson which comes
from $\phi_1^0$. For $\alpha_i =1$, $w_1 \approx 246$ GeV and $m
=100$ GeV, $Y_q \gtrsim 4.3$ can be obtained from Eq.(\ref{msigma})
when we take $2m/m_h =0.8$ and consider all quark final states but
top quark. At this moment, we must consider the light Higgs $h$
contribution to the direct dark matter detection experiments. The
WIMP-nucleon cross section by exchanging $h$ is given by
\begin{eqnarray}
\sigma_n = \frac{M^2(n)}{2 \pi} \left (\frac{\alpha_i}{m
m_h^2}\right)^2 f^2 M_n^2
\end{eqnarray}
where $f \sim 0.02 Y_q w_1/m_u$ \cite{DM}.  Using the above
parameter setting, we may derive $\sigma_n \gtrsim 6.1 \times
10^{-35} \; {\rm cm}^2$, which is far more than the $Z_1$ exchanging
case of Eq.(13). The more larger $Y_q$ is, the more larger
$\sigma_n$ is. Therefore, we can not give the desired relic number
density through increasing the Yukawa couplings.

Now we focus on the resonance case. For the $H_2^0$ exchanging case,
the results in Table \ref{Resonant} can be increased by about 5
times because of $v_R \approx 2$ TeV. On the other hand, more final
states would generally increase the partial width
$\Gamma_{H_2^0\rightarrow HH}$. Namely the case of more final states
is equivalent to enhancing $\alpha_1$, which does no good for larger
annihilation cross section as shown in Table \ref{Resonant}. For the
$h^0$ case, we may obtain the same conclusion as the one Higgs
bi-doublet case.

\section{Summary and Comments}

In the Left-Right symmetric model with one Higgs bi-doublet, we have
demonstrated that the cold dark matter constraints  should be
considered in a specific scenario in which the so-called VEV-seesaw
problem can be naturally solved. In such a scenario, we find that
$\delta_L^0$ and ${\delta_L^0}^*$ are two degenerate and stable
particles. To avoid the conflict with the direct dark matter
detection experiments, we obtain the relic number density
$n_{\delta_L^0} \leq 4.8 \times 10^{-14}$, which implies that the
two particles can't dominate all the dark matter. Subsequently, the
lower bounds  $m \sigma_0 \geq 0.13 \; {\rm GeV}^{-1}$ and $m
\sigma_0 \geq 9.8 \; {\rm GeV}^{-1}$ have been derived for the
$s$-wave annihilation and the $p$-wave annihilation, respectively.
In this paper, we examine whether our scenario can provide very
large annihilation cross sections so as to give the desired relic
abundance. We analyze in detail  four classes of annihilation
processes: $\delta_L^0 {\delta_L^0}^* \rightarrow f \bar{f}$,
$\delta_L^0 {\delta_L^0}^* \rightarrow V V$, $\delta_L^0
{\delta_L^0}^* \rightarrow H H$ and $\delta_L^0 {\delta_L^0}^*
\rightarrow V H$. However, our analysis shows that this scenario
fails to suppress the relic number density of $\delta_L^0$ and
${\delta_L^0}^*$ except for the resonance case \footnote{It needs to
be mentioned that $\delta_L^0$ and ${\delta_L^0}^*$ may be the
candidate of the cold dark matter when we use the MOND to explain
the rotation curves of disk galaxies.}. For the $h^0$ resonance
case, we obtain $\Omega_{\delta_L^0} h^2 \lesssim 6.3 \times
10^{-7}$, which is far less than the total dark matter density
$\Omega_{DM} h^2 = 0.111 \pm 0.006$. Finally, we discuss the two
Higgs bi-doublet model from the following three aspects: (1) $v_R
\approx 2$ TeV; (2) more final states; (3) large Yukawa couplings.
It turns out that our previous conclusions can be generalized to the
two Higgs bi-doublet model.

In recent years, several authors have shown  that it is far from
natural for the minimal LR model to generate spontaneous CP
violation with natural-sized Higgs potential parameters
\cite{Mohapatra,VEV,Barenboim}.  It is of importance for us to
comment on some more general LR models with one Higgs bi-doublet
\cite{Langacker,Bernabeu,Pos, Frere, Kie, Ji}. The differences
mainly come from the complexity of Higgs potential parameter
$\alpha_2$ and Yukawa couplings. We stress that our conclusion in
Sec V could be generalized to these more general cases without any
dramatic alternation because the gauge and Higgs sectors are
basically the same.

\acknowledgments{This work was supported by the National Nature
Science Foundation of China (NSFC) under the grant 10475105 and
10491306. W. L. Guo is supported by the China Postdoctoral Science
Foundation and the K. C. Wong Education Foundation (Hong Kong).}

\appendix

\section{scalar and scalar-gauge trilinear and quartic Couplings}\label{Appendix.A}

We intend to calculate the cross section to the leading order for
each process of dark matter annihilation. We first work in the
framework of simple left-right symmetric model with one Higgs
bi-doublet and one pair of LR triplets. To simplify our calculation,
we take the decoupling limit in which $\kappa^2/v_R^2\simeq 0$ where
$\kappa^2=|\kappa_1|^2+|\kappa_2|^2$ denotes the EWSB scale. The
VEVs of the Higgs bi-doublet are required to satisfy the low energy
phenomenology constraint $\kappa_2/\kappa_1\leq m_b/m_t$, which may
produce correct quark masses, small quark mixing angles and the
suppression of flavor-changing neutral currents \cite{Ecker,J.-M.
Frere,M.Raidal,Ball}. For simplicity, we take $\kappa_2\simeq 0$
which is a reasonable approximation at the leading order  since
$\kappa_2/\kappa_1$ is now around $10^{-2}$. Actually the limit
$\kappa_2\rightarrow 0$ brings additional advantage that the vacuum
CP phase  $\theta_2$ could be taken zero safely without hampering
the estimation. These approximations could largely simplify our
calculation.

\begin{table}
\begin{center}
\begin{tabular}{|c|l||c|c||c|l|}
 \hline Interaction & Coupling / $v_R$ & Interaction & Coupling / $\kappa$ & Interaction & Coupling  \\ \hline
 $\delta_L^0 {\delta_L^{0}}^* H_2^0$ & $\rho_3$ & $\delta_L^0 {\delta_L^{0}}^* h^0$ & $\alpha_1$ & $\delta_L^0 {\delta_L^{0}}^* h^0h^0$ & $\alpha_1$\\
 $\delta_L^0\delta_L^{--}\delta_R^{++}$ & $2\sqrt{2}\rho_4$ & $\delta_L^0 {\delta_L^{0}}^* H_1^0$ & $2\alpha_2$ &$\delta_L^0 {\delta_L^{0}}^* h^0H_1^0$& $2\alpha_2$ \\
 $H_2^0h^0h^0$ & $\alpha_1$ & $h^0h^0h^0$ & $6\lambda_1$ &  $\delta_L^0 {\delta_L^{0}}^* H_1^0H_1^0$ & $\alpha_1+\alpha_3$ \\
 $H_2^0H_1^0h^0$ & $2\alpha_2$ &$H_1^0h^0h^0$ & $6\lambda_4$&  $\delta_L^0 {\delta_L^{0}}^* A_1^0A_1^0$ & $\alpha_1+\alpha_3$ \\
 $H_2^0H_1^0H_1^0$ & $\alpha_1+\alpha_3$ & $H_1^0H_1^0h^0$ & $2\tilde\lambda$  & $\delta_L^0 {\delta_L^{0}}^* H_1^+H_1^-$ & $\alpha_1$ \\
 $H_2^0A_1^0A_1^0$ & $\alpha_1+\alpha_3$ & $H_1^0H_1^0H_1^0$ & $6\lambda_4 $ & $\delta_L^0 {\delta_L^{0}}^* H_2^0H_2^0$ & $2\rho_3$\\
 $H_2^0H_1^+H_1^-$ & $\alpha_1+\alpha_3$ & $H_1^0A_1^0H_1^0$ & $2\lambda_4$ & $\delta_L^0 {\delta_L^{0}}^* \delta_R^{++}\delta_R^{--}$ & $\rho_3$ \\
 $H_2^0H_2^0H_2^0$ & $6\rho_1$ & $h^0A_1^0A_1^0$ & $2\tilde\lambda'$ & &\\
 $H_2^0\delta_R^{++}\delta_R^{--}$ & $2(\rho_1+2\rho_2)$ & $H_2^0H_2^0h^0$ & $\alpha_1$ & & \\
 & & $H_2^0H_2^0H_1^0$ & $2\alpha_2$ & & \\\hline
\end{tabular}
\end{center}
\caption{The relevant trilinear and quartic scalar couplings, where
the dimensional trilinear couplings with different scales $v_R$ and
$\kappa$ are separated shown separately in two columns.
$\tilde\lambda=\lambda_1+4\lambda_2+2\lambda_3$ and
$\tilde\lambda'=\lambda_1-4\lambda_2+2\lambda_3$ }
\label{triquarcpl}
\end{table}

The relevant scalar trilinear couplings and quartic couplings under
the unitary gauge are shown in Table \ref{triquarcpl}. Here we write
out the scalar-gauge interactions:
\begin{eqnarray}
\mathcal{L}_{\delta_L^0 {\delta_L^{0}}^* VV}&=&\delta_L^0 {\delta_L^{0}}^* (gW_{3L}-g'B)^2+g^2\delta_L^0 {\delta_L^{0}}^* W_L^+W_L^-\label{DLVV};\\
\mathcal{L}_{\delta_L^0\delta_LV}&=&-ig(\delta_L^0\partial\delta_L^--\delta_L^-\partial\delta_L^0)W_L^++i\delta_0\partial {\delta_L^{0}}^* (gW_{3L}-g'B)+h.c. ;\\
\mathcal{L}_{HVV}&=&g^2v_R\left\{H_2^0[(gW_{3R}-g'B)(gW_{3R}-g'B)+W_R^+W_R^-]+(-\frac{1}{\sqrt{2}}\delta_R^{--}W_R^+W_R^++h.c.)\right\}\nonumber\\
&+&g^2k\bigg\{\frac{1}{2}H_1^{-}(W_{3R}W_L^+-W_{3L}W_R^+)-\frac{1}{2}(H_1^0+iA_1^0)W_L^-W_R^++h.c.\nonumber\\
&+&\frac{1}{4}h^0[(W_{3L}-W_{3R})(W_{3L}-W_{3R})+2(W_L^+W_L^-+W_R^+W_R^-)]\bigg\}\label{HVV};\\
\mathcal{L}_{HHV}&=&\frac{ig}{2}H_1^-\partial
H_1^+(W_{3L}+W_{3R})-i\delta_R^{++}\partial\delta_R^{--}(gW_{3R}+g'B)
-ig(\delta_R^{--}\partial\delta_R^+-\delta_R^+\partial\delta_R^{--})W_R^+\nonumber\\
&+&\frac{ig}{2}\left[(H_1^-\partial H_1^0-H_1^0\partial H_1^-)W_L^++(h^0\partial H_1^--H_1^-\partial h^0)W_R^+\right]\nonumber\\
&+&\frac{g}{2}\left[(A_1^0\partial H_1^--H_1^-\partial A_1^0)W_R^++h.c.\right.\nonumber\\
&+&\left.(H_1^0\partial A_1^0-A_1^0\partial
H_1^0)(W_{3L}-W_{3R})\right]\label{HHV},
\end{eqnarray}
where the connection between weak eigenstates $(W_{3L},W_{3R},B)$
and physical states $(Z_1,Z_2,A)$ are demonstrated by the following
orthogonal transformation at the leading order:
\begin{eqnarray}
\left(
  \begin{array}{c}
    W_{3L} \\
    W_{3R} \\
    B \\
  \end{array}
\right) &=&\left(
                   \begin{array}{ccc}
                     c_W & 0 & s_W \\
                     -s_W t_W & \sqrt{c_{2W}}\sec_W & s_W \\
                     -\sqrt{c_{2W}}t_W & -t_W & \sqrt{c_{2W}} \\
                   \end{array}
                 \right)
\left(
  \begin{array}{c}
    Z_1 \\
    Z_2 \\
    A \\
  \end{array}
\right) \; .
\end{eqnarray}
The $SU(2)_{L,R}$ gauge coupling $g$ and $U(1)_{B-L}$ coupling $g'$
are related to $U(1)_{EM}$ gauge coupling $e$:
\begin{eqnarray}
  g=\frac{e}{\sin\theta_W},\hspace{1cm}g'=\frac{e}{\sqrt{\cos2\theta_W}}
  \;.
\end{eqnarray}
Here our conventions are the same as those in Ref. \cite{DUKA}.

\end{document}